\documentclass[runningheads]{llncs}
\usepackage{graphicx}
\usepackage[misc]{ifsym} 
\usepackage{subcaption}
\begin{document}
\title{Multi-Agent Programming Contest 2019\\FIT BUT Team solution}

%
%
\author{Vaclav Uhlir(\Letter) \orcidID{0000-0002-5012-6045} \and
Frantisek Zboril \orcidID{0000-0001-7861-8220} \and
Frantisek Vidensky \orcidID{0000-0003-1808-441X}}
\authorrunning{V. Uhlir et al.}
%
\institute{Department of Intelligent Systems,\\
Faculty of Information Technology,\\
Brno University of Technology, Czech Republic\\
\email{iuhlir@fit.vutbr.cz, zborilf@fit.vutbr.cz, ividensky@fit.vutbr.cz}\\
\url{https://www.fit.vut.cz/.en}}
\maketitle			 
\begin{abstract}
During our participation in MAPC 2019, we have developed two multi-agent systems that have been designed specifically for this competition. The first of the systems is a proactive system that works with pre-specified scenarios and tasks agents with generated goals designed for individual agents according to assigned role. 
The second system is designed as more reactive and employs layered architecture with highly dynamic behaviour, where agents select their own action based on their perception of usefulness of said action.

\end{abstract}

\section{Introduction}

 This paper describes our first participation in Multi-Agent Programming Contest 2019 (MAPC). The main motivation for our participation was to compare our skill in implementing logic in multi-agent systems and try various approaches. Specifically, this text will describe two systems which we implemented for this contest -- where the first one, named ``deSouches'' (chapter \ref{deSouches}), was used for qualification and second one, named ``FIT BUT'' (chapter \ref{fitbut}), was used for the main competition.
 
\subsection{MAPC 2019 Environment}

This year, the assignment of the MAPC competition follows the story of robots that have to assemble and deliver specific structures from various blocks. For almost a decade, robotic agents have been on Mars, then on Earth, and now they are back on Mars and have to deal with an unstable volcanic environment \cite{mapc2019}.

From the classical point of view \cite{refrus} we may classify such environment as non-accessible, discrete, dynamic, non-deterministic, sequential and social. 
A non-accessible and partially observable environment provides agents with limited object visibility - where agents have limited vision to certain distance. This distance is fortunately known at the runtime and also unchanging and same for all agents during whole simulation. Agents also have limited recognition capabilities allowing them to recognize friend or foe, but not other characteristics (i.e. name or id of the seen agent). On the other hand, agents have no limitation on communication. They are able to communicate freely and thus use shared knowledge allowing them to construct and manage shared map and use it to synchronize their actions.

\subsection{DeSouches and FIT BUT Team at MAPC 2019}

Two systems that we developed for the MAPC 2019 differs in approach and aim. The first one, ``deSouches'', was a multi-agent system which proactively performed specific scenarios, and we used it for qualification. Then, thanks to the extended time between qualification and the competition, we modified the first one to a more reactive form and because the second system was performing slightly better than the first one, we used the latter in the main contest. The main goal of the competition is earning points and achieving a higher score than opponent in the current competition round. Points are earned by completing and submitting structures from found or requested special blocks scattered around the map thus requiring agents to search, assemble and deliver. The contest environment is generated at the start of the simulation and unknown to the agents and also dynamic through environment changing events. The environment is represented by a grid world containing obstacles, dispensers (places where blocks are issued) and goal marks - places where agents have to place assembled shapes.

Due to the design of competition simulation -- agents are required to be able to actively cooperate with each other in the process of assembling structures. As a minimum for agents trying to join two blocks is active assistance of another agent from the same team. Other tasks, such as searching, delivering, clearing terrain or even targeted malicious action do not require team cooperation/synchronization but may provide a clear advantage when implemented.

\section{deSouches Multi-agent System}\label{deSouches}


The first problem addressed was the question of which architecture, language, implementation system, and multi-agent methods would be used for successful implementation of desired methods. We evaluated a number of solutions that were possibly suitable both for the implementation of a rational agent and for the use of multi-agent methodologies for teamwork or conflict resolution. Among the most important was JASON \cite{ref2}, which interprets the language AgentSpeak (L) \cite{ref1} and, together with 2APL \cite{ref4}, is probably the best-known representative of systems that work on the basis of the BDI paradigm \cite{ref3}. Extending JASON to JaCaMo \cite{ref6} is a workable way to create multi-agent assemblies that pursue common goals and this would certainly be appropriate for creating systems for MAPC as this was successfully implemented by authors over the past seasons. But as this year was a completely new scenario, we decided to try to use our experience with agents and a multi-agent system and create our own multi-agent system.

We decided to do the brand-new system that would fit the contest scenario and to implement it in JAVA. The reasons were rather intuitive. Beside an opportunity to try to make an agent system from scratch we had doubts about performance of today's agent systems in such a highly dynamic environment. We realized that it would be better to spent more time in implementation but to have the performance of the system under our own control. Finally we saw that this enabled us to change architecture of the system in quite short time to more reactive allowing improved behaviour of the agents, as we found out by comparison of the original and latter version.

The first architecture was made for the purposes of qualification. It was inspired by the BDI solutions mentioned above, but this system was more proactive with persistent goals rather than regular BDI system. In the following lines, we will describe this first system and in chapter \ref{fitbut} we will present the modified system that we used during the main content.

\subsection{General deSouches and his soldiers}
It is clear that the competition assignment was created, among other things, to verify the social ability of a multi-agent population in a dynamic and non-deterministic environment. It addressed the ability of agents to make a coalition and jointly follow a goal. This is still an ongoing issue in multi-agent systems and each team has to address this problem.

The first multi-agent system we named after a general who successfully commanded Brno's defence force against the siege of the Swedish army during the Thirty Years' War in 1645. We formed a group of agents in a very simple hierarchy where one agent – deSouches assumed commanding of others. The commands were in the form of scenarios that the soldier had to adopt. The soldiers then followed the scenario as they got goals that they should complete. Soldiers also informed deSouches when they successfully finished the scenario or when they failed to complete the scenario or simply that they needed a job.

\subsection{Agent level architecture}
Originally, our first agent architecture was designed to be one that works with a set of intentions. The intentions were created for a persistent goal and the agents build a hierarchy of procedural goals \cite{ref5} and corresponding plans that should lead to the main intention goal achievement. But in the process of development, we ended up with a system where agents could contain only one intention and also the intention contained every time only the main goal. Thus the agent was driven by procedural goals that were represented by a goal class that includes methods for making a plan for the goal and processing of one action of an actual plan. Agent’s beliefs about the environment consisted of a map of the environment and other information that it may get from its percepts – agent’s own name, team name, energy and blocks in visible vicinity. The last contains the agent and all the attached blocks.

A plan is made from actions available to the agents according to the contest specification. An agent may traverse the environment, make clearing actions, attach, detach and connect blocks, rotate and submit tasks. Essential for our agent was that for selection of goals it could use A* search algorithm. In this case, it was slightly modified with the restriction of the number of iteration it could make. It was often impossible in a given map to get from one place to another, especially when the agent’s body contained one or more blocks. Then the A* was unsuccessful and the planning for a goal failed.

Interpretation cycle was implemented as commonly expected -- that is as it consisted of reasoning and execution parts --. During the reasoning part, the agent firstly processed its percepts then evaluated feedback from previously executed action (as it could have failed in the environment) and sent actual tasks proposed by the system to deSouches. Then agent took its intention, respectively the goal within the intention, reconsidered the plan for the goal and then executed one action of the plan.

\subsection{Goals}
There were about ten different goals that were specifically implemented for our agents. The agent has a choice of adopting a goal of traversal to a specific position. Or the goal of traversal where the agent would walk randomly to some specified distance. Several goals were related to the spacial manipulation of the blocks, block retrieval or to attachment and detachment of the blocks and finally to submitting blocks in a given direction. The last-mentioned goals were quite simple and the plans contained only a few actions, usually to rotate to the given direction and to attach, detach or submit a block.

\subsection{Synchronization of agent groups}\label{firstsynchro}
Because the agents do not know their absolute position on the environment grid they are also unable to deduct their relative positions to each other. If one agent discovers an obstacle, dispenser or block, other agents are not able to simply calculate their relative position to discovered object. Thus synchronization of agent positions was essential for successful behaviour of our agents. Our key idea for such synchronization is that when there is only one pair of agents that one agent sees friendly agent relatively to it on $dx$, $dy$ and another agent sees friendly agent on $-dx$, $-dy$, then these two agents see one another and may synchronize. This will be described in more details further in the section \ref{synchro} in part of our second multi-agent system. Furthermore, we form agent groups that contain synchronized agents. At the beginning, we start with $n$ groups for $n$ agents. When there is a pair of agents that may be synchronized and these two agents are from different groups then we will join these groups together and merge their maps with respect to the shift vector between the pair of just synchronized agents. Consequently, we build up larger and larger groups and we can see that at most after $n-1$ synchronizations there remains only one group containing all the agents and then our whole team is fully synchronized.

\subsection{Scenarios}
Scenarios are the main source of the behaviour of deSouches's agents in the game. Scenarios are in fact finite-state automatons by which the goals are assigned to agents. There is one automaton for every role in the scenario and as the goals are successfully or even unsuccessfully performed the automaton may change its state and assign another goal. When all the agents reach the final states then the scenario is completed and deSouches is informed. It may also happen that the scenario cannot be completed, for example when the deadline for a task is over and then deSouches is again informed with this fact.

There are three basic kinds of scenarios in our system. In the beginning, the agents need to explore the environment and get synchronized. After they are fully synchronized they will either work on a task or try to clear an area important for the team. A subgroup of agents may also be chosen for a task before the multiagent population is fully synchronized. Such a group must have number of agents equal or greater than the number of block needed and also beliefs where particular depots for needed types of blocks are. When deSouches creates a group of agent for a task, then such agents must try to complete the task. Before we will introduce the scenario we have to discuss the problem of synchronisation of agents in the environment.

\subsubsection{Walk and synchronize scenario} is the starting point for every agent. Exploring the map is achieved by soldiers that walk randomly in order to explore the map of the environment and also to synchronize the agents that are meeting while traversing the unknown environment. We will describe the synchronization later in this text. An agent that is supposed to walk randomly has to select a direction and with a distance in pre-specified range it chooses a point where it should go. Then it computes path using $A*$ algorithm, which is modified in such a way that the number of iterations is limited by a specific number (2500 proved to be adequate). This guarantees that it ends in sufficient time and also that it ends even when there does not exist a path to the destination. In either case, the agent follows the computed path returned by the algorithm. After it encounters an unexpected obstacle or finishes successfully given steps, the scenario is over and the agent sends corresponding (success/fail) message to deSouches. In the next step, deSouches sends back a new order to repeat this scenario in a new variant again or assigns the following search and destroy scenario.

\subsubsection{Search and destroy scenario} is ordered to agents after there is only one group of agents, this means that all of the agents are synchronized. In this case, the agents should be aware of where are the obstacles that need to be destroyed and if they can successfully reach them, then they try to destroy them by the clear actions. When they cannot get to the chosen obstacle or they do not have enough power to destroy it they report a failure to deSouches. Otherwise, after successful completion of this scenario, they report success to the general deSouches. 

This scenario offers several alternative strategies on how to choose an obstacle for elimination. We considered that the agent could try to remove obstacles around the goal areas or dispensers. But in the resulting algorithm in the finished development stage of deSouches architecture the agents just take the nearest obstacles they can see and clear them.

\subsubsection{Blocks scenarios} are derived from the same class and are predefined for two, three and four blocks. For each of them, there is one master agent and one, two or three lieutenants. DeSouches, when discovers a possibility of fulfilling of a selected task and has available ``mastergroup'' that has more free agents than is required number for two, three respective four-block scenario -- meaning at least two, three or four agents, initializes such scenario. Through this initialization, deSouches selects suitable agents from the mastergroup.


In case when there is no mastergroup yet, deSouches looks if there is a group that knows the position of all demanded dispenser types for the given task and also if there are enough number of unemployed agents in the group. If there is such a group, deSouches employs agents from this group for the tasks and the group also becomes a mastergroup. Once labelled as mastergroup the group remains mastergroup for the rest of the simulation run. During any further synchronizations, any other group will be synchronized with this group.

\begin{figure}[ht]
\includegraphics[width=\textwidth]{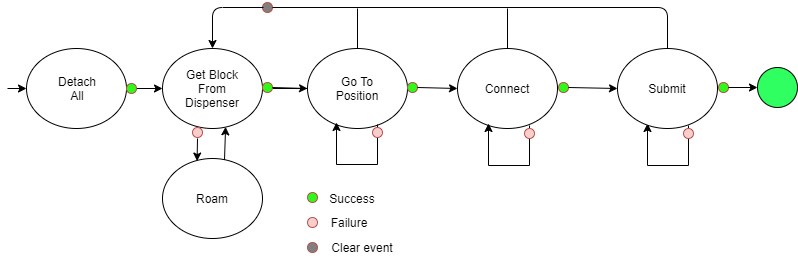}
\caption{Automaton for deSouches blocks scenario.} \label{des_flow}
\end{figure}

The agents have individual roles in this scenario. For every block scenarios, one agent is set as commander and one as the first lieutenant. For a three-block scenario, there is also the second lieutenant and for four block scenario, there is another -- the third lieutenant. Commander would be finally responsible for submitting the task and we present the corresponding automaton for this role in this scenario in figure \ref{des_flow}.

For lieutenants, the diagram is mostly the same except for the Submit goal -- which is in this case omitted.

Each participant knows which type of block it should obtain and how to connect them. More specifically each agent in his role knows with which type and on what position in or nearby a selected goal cell he should stand (only the commander must stand on a goal cell). Agents also know the direction of their block must face before they will try to connect them together.

The blocks scenarios will use one of in total 42 plans that the agents are able to make in our system. This allows agents to plan and execute two, three and four block shapes assembles and submit scenarios.

There are three plans for shapes consisting of two blocks, ten plans for three-block shapes and the remaining twenty nine plans are for structures that are created from four blocks. In this agent implementation, we programmed them by enumeration instead of developing an algorithm that would generate such plans dynamically, as was originally intended at the very beginning of development.

Every agent in the team must know which type of block it should acquire and which position around the goal area it must take before it tries to connect the block in a given direction. If an agent manages to get the proper block then it tries to get to its goal position and then to connect the block with its teammates. If any plan goals fail then it tries to reach it again except the ‘go to dispenser’ goal where the agents try to perform a random traversal first before trying to achieve the agents intended original goal. Also if a clear event is fully performed on agent and results in loss of the block -- agents must restart the scenario from the ‘dispenser’ goal again. When all the blocks are properly connected, all the agents except the commander detach their blocks and then the commander submits the task.

If every agent is able to complete successfully their automatons then the scenario should be successfully completed and the playing team should receive points for simulation task completion. The scenario may also fail when the task deadline lapses and in such case, general deSouches disbands the team and gives new orders to the agents –- to perform another scenario.

\section{FIT BUT System}\label{fitbut}

The second system that we used for the final contest is named simply FIT BUT. In some parts, it overlaps with deSouches system (as it was built on its bases), but it is more reactive using a hierarchically layered model of behaviours and it is rather similar to Subsumption architecture \cite{ref7} than the previous system was. After we compared both systems we found out that more reactivity instead of proactivity gives in such a dynamic environment as the MAPC 2019 better outcomes and for this reason we used FIT BUT for the main competition.

\subsection{Design}

\begin{figure}[ht]
\includegraphics[width=\textwidth]{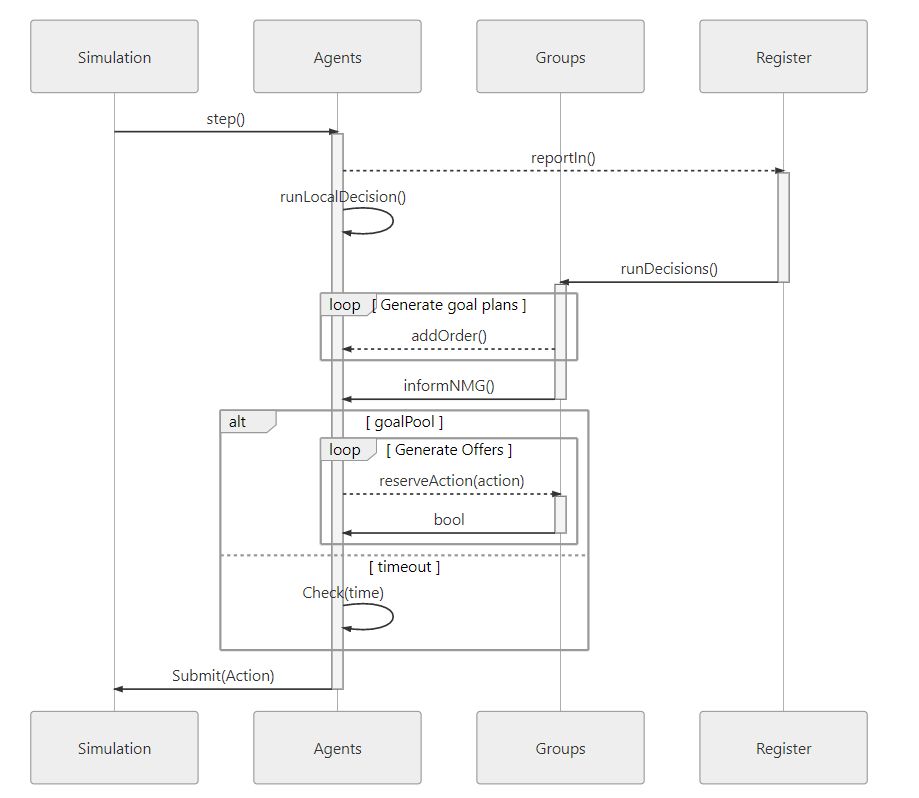}
\caption{Overview of one step in FIT BUT system.} \label{flow}
\end{figure}

Basic system flow can be seen in figure \ref{flow} and consists of three levels of granularity. The logical structure of the system is again hierarchical and agent units, supplied by system percepts are organized into population groups. Each such a group is then registered into global Register implemented by singleton class.

At the start of every step, agents evaluate their percepts and then contact Register class. After every agent successfully registers completion of percepts evaluation, Register -- if needed -- assigns agents to their new group (if new relative positions are confirmed) and contacts all groups and triggers groups calculation of agent options. Groups are informing agents upon every successful action option generation and it is left to the agents to select their future action, but it still has to be confirmed by the group reservation system. Upon successful confirmation, the agent in question returns desired action to simulation or if a reservation was unsuccessful has to find different action which will be confirmed by the group reservation system.

Every plan generated in this system contains a full set of actions following a purpose to the specified goal. But at every step in this system the old plan is replaced by a new plan influenced by environmental changes, thus only one (the first) action from every plan is ever used. This ensures fully reactive behavior and also allows clear tools for behavioral analysis.

\subsection{Synchronization}\label{synchro}
The key features defining agent synchronization are agent \textit{visibility}, \textit{communication capabilities} and \textit{certainty}. We discussed the environment briefly at the beginning of this text. The main issue was that agents do not know the disposition of their environment or starting position in relation to global map or to each other -- but all agents have same orientation sense and most actions produce new determined position -- either successfully executing (non)movement action with a new expected position, or getting action error with position unchanged. An Agent can reach (in his view) unexpected position by being dragged by another friendly agent in case where two agents are connected via blocks to the same total structure. This behavior can be eliminated by carefully monitoring agents block attachments and restricting movement actions while more agents are connected to the same structure.

As mentioned at the beginning of this text, the agents are not able to recognize other agents in their vision beyond friend or foe. Fortunately, this is not the case in communication and messages can be signed and trusted allowing agents to compare their vision results. Due to the non-observable and dynamic environment using landscape recognition does not produce certain results with possible exception based on the border mapping by movement actions error logging. This process may be lengthy and in a highly dynamic environment nearly impossible as borders may not be reachable in meaningful time.
 
\begin{figure}[ht]
\centering
\includegraphics[width=0.5\textwidth]{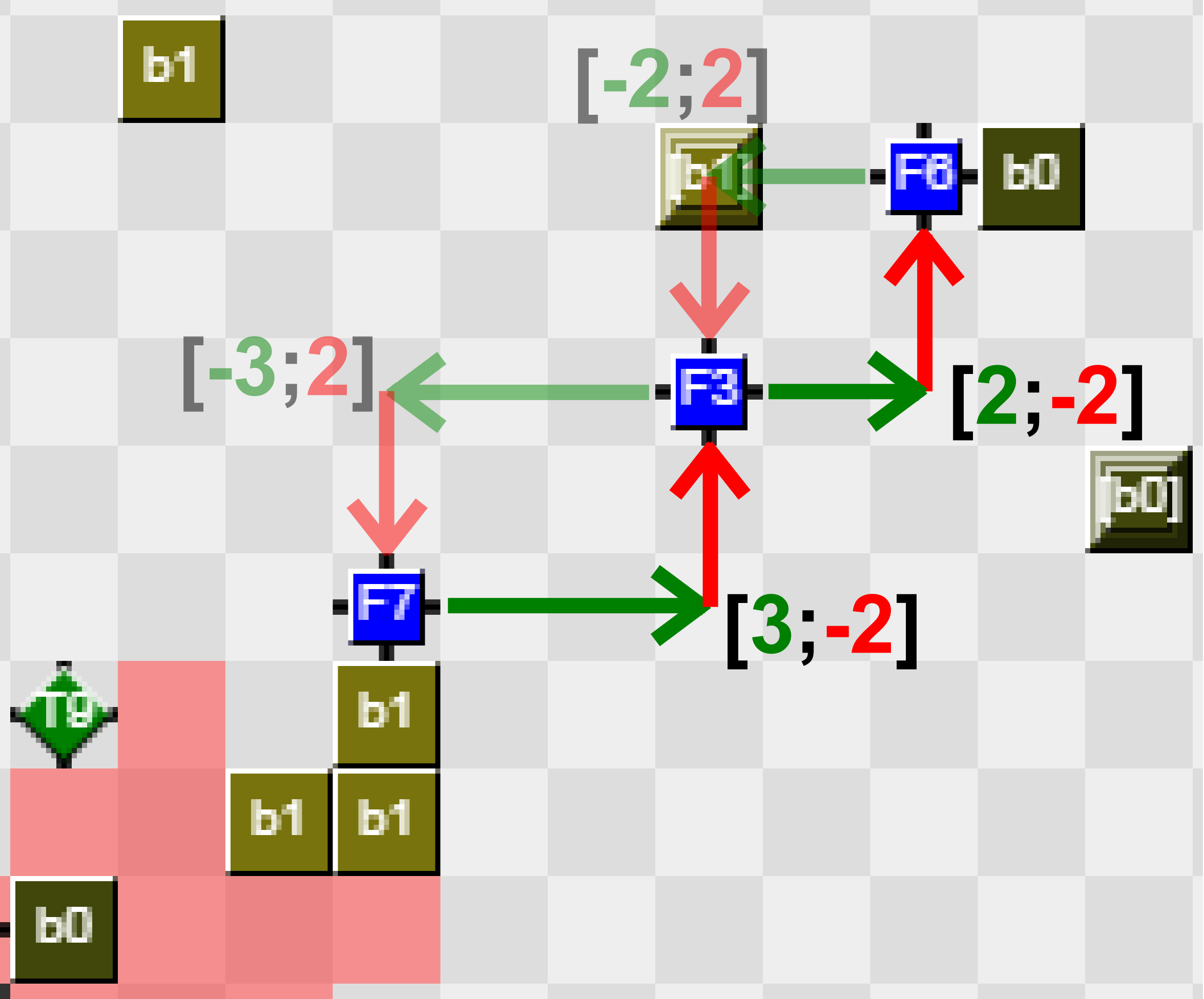}
\caption{Example of using distance difference for agent recognition.} \label{sync}
\end{figure}

Based on these condition FIT BUT agents use position synchronization by implementing two-way confirmation solely based on seeing each other. This can be trusted as synchronization procedure in case of exactly two agents seeing another agent at the same distance but in opposite direction and if no other agents detect any other agent in the same distance and direction (Fig. \ref{sync}) these two agents can be certain of their current relative position difference and with successful tracking of their own action they can determine their position relation for the rest of the simulation. 

\subsection{FIT BUT Agent Reasoning Cycle}
FIT BUT system works on bases of options calculation that the agents try to discover during their interpretation cycle. Due to the limitations of resources \cite{ref9}, especially time constraints, not every option may be discovered. Discovered options are formed into the possible plans which are sorted based on priorities and resources availability with the aim of maximizing points. In every agent step, only the first action of the plan for the option with the highest priority is executed and in the following step, the agent computes its options again. Thanks to this approach, the agent is very reactive and careful because it reconsiders its behaviour in every cycle. Below we will specify these phases in more detail.

\subsection{Options analysis}\label{opcalc}
\begin{figure}[ht]
\centering
\includegraphics[width=0.65\textwidth]{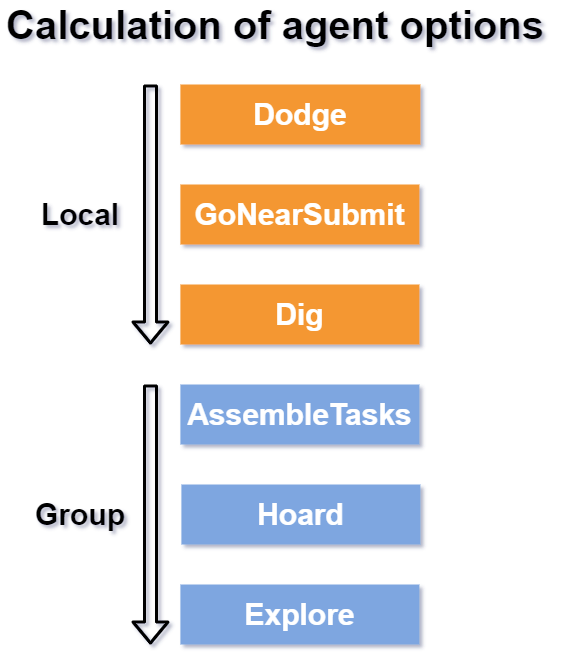}
\caption{Options discovery order.} \label{calcprio}
\end{figure}

After agents evaluate their perception both local and group decisions calculations are triggered simultaneously and in order as seen in Fig. \ref{calcprio}. While calculating options, the watchdog is checking time and may terminate calculations if calculation should exceed current step timeout.

\subsubsection{Local decision} calculations are executed immediately in order of \textbf{Dodge}, \textbf{GoNearSubmit} and \textbf{Dig}, where:
\begin{itemize}
	\item \textbf{Dodge} checks if any clear action is expected on agents (or his body part) position and if needed plots shortest escape course out of clear action influence area.
	\item \textbf{GoNearSubmit} finds nearest (currently seen or remembered) goal position and calculates a path to it.
	\item \textbf{Dig} plots path to the closest terrain and if the terrain block is in range triggers clear terrain action for such block.
\end{itemize}
The first locally analysed option \textbf{Dodge} is intended to protect agents integrity especially when carrying blocks. This action takes precedence over most other actions and has priority in the reservation system. Also when this action is present in possible actions stack, low priority actions may be excluded from the calculation. 
Remaining local actions \textbf{GoNearSubmit} and \textbf{Dig} are intended as fallback on group coordination problems and timeouts. Action \textbf{GoNearSubmit} is intended to bring agents close to submit areas to decrease their collective distance difference or enable discovery of other friendly agents. Remaining action \textbf{Dig} is executed as a last resort if no other action and thus no path to a goal area is found. This is intended for agents stuck inside terrain without known access to useful space.

\subsubsection{Group decision options} calculation is started when the last of the agents reports finished evaluating perceptions. As the calculation of group options may be very resource-intensive -- it is expected that not all of the options will be calculated and agents with priority plans already assigned are excluded from further options calculation. Group options discovery is performed in order of \textbf{AssembleTasks}, \textbf{Hoard} and \textbf{Explore}, where:
\begin{itemize}
	\item \textbf{AssembleTasks} searches for blocks and paths for active tasks (detailed in Section \ref{asstasks}).
	\item \textbf{Hoard} fetches and attaches blocks to the agent.
	\item \textbf{Explore} roams agent into unknown or longest unseen parts of the environment.
\end{itemize}

The first considered option: \textbf{AssembleTasks} is the one most resource-intensive, but most important in scoring points. This calculation is more resource heavy more block agents have hoarded and more space have they explored, so starving \textbf{Hoard} and \textbf{Explore} calculation is not as critical in such case and in case of no blocks hoarded or goals explored this action is entirely skipped.	

\textbf{Hoard} action is intended for acquiring of blocks (either retrieving already existing blocks in the environment or obtaining block from the dispenser) and attaching them to the agent body.

The last group goal \textbf{Explore} aims at discovering new dispensers, blocks, goals and also other agents. Agents in cooperation group divide the unknown area into section and each agent tries to discover assigned part (usually closest to its position). If no unknown areas are accessible agents will perform a check of areas that are longest unseen (this is aimed to monitor terrain changes and find possible previously undiscovered areas).

\subsection{Plan selection}
Every agent is selecting its plan for execution in the current step from the available plans created for the options discovered as described above in section \ref{opcalc}. Every option produces a plan in the form of an action stack where this actions correspond to the actions that are available for the agent in MAPC. Names of the plans are the same as the names of corresponding options with exception of plans \textit{GoConnect} and \textit{GoSubmit} which are created for the \textbf{AssembleTasks} option and the plan \textit{Roam} that is created for the \textbf{Explore} option. 

\begin{figure}[ht]
\centering
\includegraphics[width=0.7\textwidth]{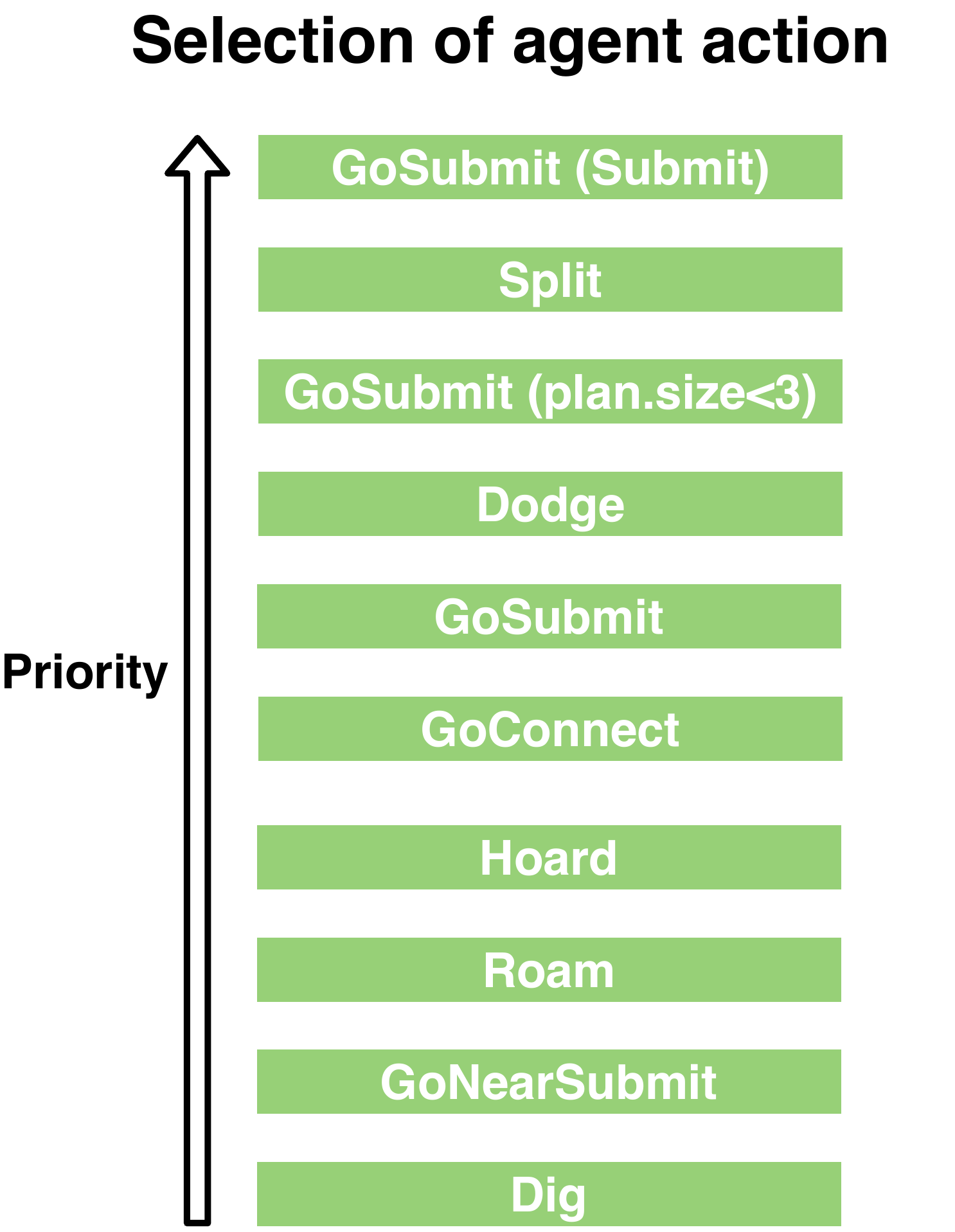}
\caption{Priority order of plan selection in which agent tries to execute selected operation.} \label{priority}
\end{figure}

Furthermore, there may be also a plan named \textit{Split} that was created immediately after a connection action is successfully executed and is persistent until agents are separated.
Plan intended for execution is selected in priority order as shown in Fig. \ref{priority}. 

Plan with the highest priority is \textit{GoSubmit} and is adopted when an agent is able to submit part of his current body as an active task requirement. If such action is not possible and the agent is still connected to another agent then the agent results in plan \textit{Split}. This should eliminate unwanted position change by agents dragging each other by other movement actions. 

Next in order is again \textit{GoSubmit}, this time with limitation on its plan size -- more precisely when the plan is less than $3$ actions in length. Such a plan is preferred to the next option --\textit{Dodge}. This is in line of reasoning that continuing of submission shall probably be finished before the clear action is performed and therefore advantageous. After plan option \textit{Dodge} follows again plan \textit{GoSubmit} this time without the additional limitations intended for execution of the rest of its variations.

If agent doesn't have task which it may submit and none of the above mentioned types of plans are available, then the agent follows with the selection in the order of \textit{GoConnect}, \textit{Hoard}, \textit{Roam}, \textit{GoNearSubmit} and \textit{Dig}. 

The reason for an agent to create new plans that will inevitably have lower priority than already created plans from previous options is to allow the agent to generate backup actions in a case where the agent may not be able to perform previously selected action with a higher priority. As mentioned above, the agent has to select a new plan in every interpretation cycle and from this plan agent will execute only the first action. To mitigate team collisions the agent has to check his action against the actions of his teammates and thus before action is executed it has to be confirmed by the system that no collision shall occur in cases of successful or even unsuccessful actions. This safety feature is performed by the action reservation system.

\subsection{Action reservation system}
\begin{figure}[ht]
\begin{subfigure}{0.49\textwidth}
\includegraphics[width=0.9\linewidth]{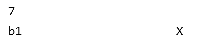}
\caption{Agent 7 reserving block request.} \label{reserv_a}
\end{subfigure}
\begin{subfigure}{0.49\textwidth}
\includegraphics[width=0.9\linewidth]{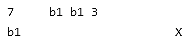}
\caption{Agent 3 reserving step to the right.} \label{reserv_b}
\end{subfigure}
\begin{subfigure}{0.49\textwidth}
\includegraphics[width=0.9\linewidth]{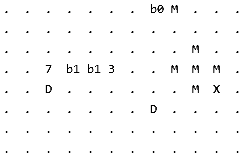}
\caption{Starting step state.} \label{reserv_c}
\end{subfigure}
\begin{subfigure}{0.49\textwidth}
\includegraphics[width=0.9\linewidth]{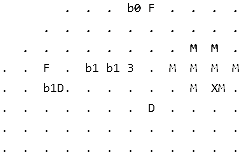}
\caption{Resulting next step.} \label{reserv_d}
\end{subfigure}
\caption{Successful reservation and execution of agents actions.}\label{reserv}
\end{figure}

As mentioned above all agents, when choosing an action for execution, must first contact reservation system (as can be seen in Fig. \ref{flow} with call \textit{``reserveAction(..)''}) which creates a map of expected result for the next step and required free spaces for movement or action operations. This map is shared for all agents in the current synchronized group. Reservation system then can approve the agent action and reserve needed cells (or blocks or other resources) or can reject their action which results in agents having to submit different action for review. Successful reservation can be seen in log files from competition on Fig. \ref{reserv} where (after successful structure joining and agents disconnecting) agent $7$ wants to request a new block from dispenser bellow (Fig. \ref{reserv_a}) and then agent $3$ requests traverse movement to the right towards his goal (Fig. \ref{reserv_b}) and thus resulting successful operations will change the environment from original step (Fig. \ref{reserv_c}) to the next step (Fig. \ref{reserv_d}).

\subsection{Task possibilities commutations}\label{asstasks}
One of the crucial parts of FIT BUT system is finding suitable structures that may be used for assembling active tasks. Assembling has to be done by the agents (which have already hoarded some blocks and are still in possession of these blocks) and have to be performed by collaboration and following simultaneous joining call. Analysis on which structures shall be joined to achieve desired shapes and types is performed for the \textit{AssembleTasks} option and is done algorithmically as described by the following steps:
\begin{enumerate}
	\item Insert every agent to connection candidates set (CCS).
	\item Evaluate currently held structures with active tasks, if any agent is already in possession of structure fitting to a requested task structure and the agent can plot a path to a goal area, then create a new \textit{Submit} plan for such agent and remove it from the CCS.
	\item Sort the structures carried by agents in order from most complete to least complete in regard to required tasks and their value.
	\item For every active task and every agent in CCS, generate combinations of agent pairs and their structures fitting targeted tasks with respect to their order from the previous step. All generated combinations have to be connected and cannot overlap on any block.
	\item Sort the generated structures by metric composed of steps needed for their assembly and reward for the corresponding active task.
	\item While there is a pair of agents in CCS that were identified as suitable pair for any connection:
	\subitem Select a generated structure with the highest metric value where both agents are in CCS.
	\subitem Create for such agents corresponding \textit{GoConnection} plans \subitem Remove these agents from CCS.
\end{enumerate}

Generating combination for every task, every agent and every block would be very resource-intensive and therefore sorting and option pool trimming is implemented on multiple points in the algorithm. 

\section{Limitations and possible improvements}
Our agents are limited in lots of aspects due to planned but unimplemented features which could improve future potential.

\subsection{Size control}
One of such features is structure size control -- tool for checking size of the future connected structure against the system limit. For example, in testing matches the maximal size of the structure was set to $10$ blocks. Let us suppose two agents with a connected block on their every side -- connecting such entities would result in a structure with $8$ blocks and $2$ agents thus totaling $10$ blocks and triggering the system limit.\\

\subsection{Hoarding}
Hoarding routine is implemented for connecting blocks to all of the sides of the agent but due to missing size control was limited to a maximum of one block, which proved useful enough for time being and other issues took precedents. Upon implementation of size control agents should benefit from faster block matching for all the tasks. The downside of enabling hoarding for more than $1$ block is the limitation of agent movement capabilities as it would not be possible to manoeuvre through tight places and either dynamic block dropping or additional terrain digging may be required for maintaining agent mobility.

It should be noted that hoarding section of code contains function \textit{isBlockInteresting(BlockType)} intended for management of block type possession of the whole group. This function was not yet properly implemented and always returns \textit{true}. While this missing functionality was fortunately not the issue during competition it poses the risk of agents not hoarding diverse enough blocks and thus limiting their capability to fulfill task (as current task possibilities are computed only from hoarded blocks and agents with hoarded wrong blocks can deadlock their assembly possibilities).

\section{Conclusion}	
Both of our solutions described in this text were able to complete a number of given tasks in the MAPC 2019 game scenarios. As was described in sections \ref{firstsynchro} and \ref{synchro} synchronization of the agents is one of the key aspects on road to success and necessity in such dynamic environment with agents having only limited vision capability and without information about their absolute position within a coordinate system. Another obstacle described and solved in the work above is problem of agents recognition -- final solution of this problem is relatively simple, but as explained in section \ref{synchro} it is efficient enough and furthermore reliable without fault. Using this method our algorithms were able to form work groups. 

The first described system named deSouches (section \ref{deSouches}), that was used for qualification, presented close multi-agent cooperation and assembling the tasks using agents from one synchronized group. Each agent from this group was assigned with retrieving and transporting one specific block for the chosen task. This block was transported to the goal area where all of the tasked agents connected them to the required shape and agent assigned as commander of the group submitted them for the specified reward. 

The second described system, with assumed team name FIT BUT (section \ref{fitbut}), that was eventually used for MAPC competition was more reactive and used agents to collect blocks as an opportunity arose. Such collected blocks where then connected to form expected shapes or their sub-parts. Such sub-parts were formed opportunistically and without a set future plan, but thanks to constant reevaluating in every step this proved to be quite functional and mainly very robust when it came to overcoming environment events and enemy agents actions.

Both of the systems had their advantages and disadvantages -- first one more constant in outcomes end with pre-planned goals proved to have higher base reliability. The second system was thanks to its reactivity far more versatile and dynamic, but with dangers of bottlenecking or encountering obscure problems in specific scenarios. Both of the systems proved their main required features in their respective competition rounds.

\section*{Acknowledgment}
This work was supported by the project IT4IXS: IT4Innovations Excellence in Science project (LQ1602).

\appendix
 \section{Team overview: short answers}

 \subsection{Participants and their background}
 \begin{description}
	\item \vskip0.5em\textbf{What was your motivation to participate in the contest?}
	\\ Our group is related to artificial agents and multi-agent systems and we wanted to compete in an international contest and test our skills.
	
	\item \vskip0.5em\textbf{What is the history of your group? (course project, thesis, $\ldots$)}
	\\ Members of our research group have been teaching artificial intelligence at our faculty for nearly 20 years. Most of the projects or thesis in our group concern the topic of artificial intelligence, multi-agent systems, soft-computing and machine learning.
	
	\item \vskip0.5em\textbf{What is your field of research? Which work therein is related?}
	\\ Vaclav Uhlir: Ecosystems involving autonomous units (mainly autonomous cars).
	\\František Zboril and Frantisek Videnky: Artificial agents and BDI agents. Frantisek Zboril's field of research is also prototyping of wireless sensor networks using mobile agents. 
	
 \end{description}

 \subsection{Statistics}
 \begin{description}
	\item \vskip0.5em\textbf{How much time did you invest in the contest (for programming, organizing your group, other)?}
	\\Something between 200 to 300 hours of programming and another 100 of planning, strategizing and managing $git$ and other development environments.
	\item \vskip0.5em\textbf{How many lines of code did you produce for your final agent team?}
	\\5531 lines of code.
	\\797 comment lines.
	\\42 active ``TODO'' in final code.
	\item \vskip0.5em\textbf{How many people were involved?}
	\\3
	\item \vskip0.5em\textbf{When did you start working on your agents?}
	\\ Aug 29, 2019 10:41am.
 \end{description}

 \subsection{Agent system details}\label{sec:strategies}
 \begin{description}
	\item \vskip0.5em\textbf{How does the team work together? (i.e. coordination, information sharing, ...) How decentralised is your approach?}\\
	Every agent has its local tasks with priority list as a fallback and if time allows, it waits for local group decision (triggered by the slowest agent in the group).
	
	\item \vskip0.5em\textbf{Do your agents make use of the following features: Planning, Learning, Organisations, Norms? If so, please elaborate briefly.}\\
	Our agents plan is only one step and agent are organized into groups as they "meet" - within these groups agents cooperate based on momentary advantage.
	
	\item \vskip0.5em\textbf{Can your agents change their behavior during runtime? If so, what triggers the changes?}\\
	Every action is dependent only on the current environment and few randomizers independent on previous steps. 
	
	\item \vskip0.5em\textbf{Did you have to make changes to the team (e.g. fix critical bugs) during the contest?}\\
	Yes, we enabled not-fully-tested beta features hoping to achieve better error handling. 
	
	\item \vskip0.5em\textbf{How did you go about debugging your system?}
	Custom logger with 5 levels of logging for every agent and bound to various subsystems. (By average every contest match produced around 1GB plain-text info.)
	
	\item \vskip0.5em\textbf{During the contest you were not allowed to watch the matches. How did you understand what your team of agents was doing? Did this understanding help you to improve your team's performance?}\\
	An overwhelming flood of error indicated network problems and resulted in agent desynchronization - limiting the system higher functions. Enabling beta features eliminated some network issues but introduced other errors.
	
	\item \vskip0.5em\textbf{Did you invest time in making your agents more robust? How?}\\
	Robustness was planned via fallback strategies -- some of them were implemented in beta features, but most was not ready for the main contest. 
 \end{description}

 \subsection{Scenario and Strategy}
 \begin{description}
	\item \vskip0.5em\textbf{What is the main strategy of your agent team?}
	\\ Aiming for closest possible highly valued target while effectively ignoring past.
	
	\item \vskip0.5em\textbf{Your agents only got local perceptions of the whole scenario. Did your agents try to build a global view of the scenario for a specific purpose? If so, describe it briefly.}
	\\ Upon successful position confirmation of any two agents, agents were assigned to workgroups synchronizing any new perception to the global map used in the search for highest achievable task completion.
 
	\item \vskip0.5em\textbf{How do your agents decide which tasks to complete?}\\
	When a task is available and all required blocks are accessible and their joining can result in the successful completion of structure and structure can be delivered to the goal. Task value selection is based on reward and the number of steps needed with a slight preference for smaller structures.
	
	\item \vskip0.5em\textbf{Do your agents form ad-hoc teams to complete a task?}\\
	Agents cooperate in team-like structures but in every step, the cooperation can be reevaluated.
	
	\item \vskip0.5em\textbf{Which aspect(s) of the scenario did you find particularly challenging?}\\
	Identification of block attachments (to each other or agents).
	
	\item \vskip0.5em\textbf{If another developer needs to integrate your techniques into their code (i.e., same programming language tools), how easy is it to make that integration work?}\\
	Definitely under average as some in-code named features are not fully complete and/or are using various temporal workarounds. 
 \end{description}

 \subsection{And the moral of it is \ldots}
 \begin{description}
	\item \vskip0.5em\textbf{What did you learn from participating in the contest?}\\
	Relatively simply looking scenario can present a far greater challenge than we expected.
	
	\item \vskip0.5em\textbf{What are the strong and weak points of your team?}\\
	Our team has expertise in multiple different languages and coding approach techniques.\\
	Every member has specialization in different programming language and techniques.
	
	\item \vskip0.5em\textbf{Where did you benefit from your chosen programming language, methodology, tools, and algorithms?}\\
	Familiarity with the used environment allowed for faster development for some team members.
	
	\item \vskip0.5em\textbf{Which problems did you encounter because of your chosen technologies?}\\
	Mainly portability issues over different operating systems and conflicting environment variables.
	
	\item \vskip0.5em\textbf{Did you encounter new problems during the contest?}
	Yes - battling environment and OS portability on large scale.
	
	\item \vskip0.5em\textbf{Did playing against other agent teams bring about new insights on your own agents?}
	Yes - mainly highlighting strength and weaknesses and opening ideas for new strategies. 
	
	\item \vskip0.5em\textbf{What would you improve (wrt. your agents) if you wanted to participate in the same contest a week from now (or next year)?}
	Error handling, network code, fallback strategies - in this order.
	
	\item \vskip0.5em\textbf{Which aspect of your team cost you the most time?}
	In the early versions of the system, we had a nasty bug that sometimes caused subsequent errors in synchronization amongst agents and problems in other systems. This was blamed on various other possible sources and caused very lengthy bug-hunting through multiple environments. 
	
	\item \vskip0.5em\textbf{What can be improved regarding the contest/scenario for next year?}
	Clarification about block connections - either changing perceptions with some sort of connection information or clear warning about the uncertainty of block connections.
	
	\item \vskip0.5em\textbf{Why did your team perform as it did? Why did the other teams perform better/worse than you did?}\\
	Our agents were running on a machine with desynchronized clock (about -3.5sec) and thus fearing timeouts, agents were submitting action prematurely with less than 0.5sec on decisions - which they were not built for and because of this the higher system planning was often not effectively used.
	\end{description}
%
%
%
%

\end{document}